\documentclass[12pt]{article}      

\title{Geometric Probabilities and Fibonacci Numbers for Maximally Random n-Qubit Quantum Information States}

\author{ Oktay K Pashaev\\Department of Mathematics\\ Izmir Institute of Technology \\ Urla-Izmir, 35430, Turkey}

\begin{document}
\newcommand{\be}{\begin{equation}}
\newcommand{\ee}{\end{equation}}
\newcommand{\bea}{\begin{eqnarray}}
\newcommand{\eea}{\end{eqnarray}}
\newcommand{\disp}{\displaystyle}
\newcommand{\la}{\langle}
\newcommand{\ra}{\rangle}

\newtheorem{thm}{Theorem}[subsection]
\newtheorem{cor}[thm]{Corollary}
\newtheorem{lem}[thm]{Lemma}
\newtheorem{prop}[thm]{Proposition}
\newtheorem{definition}[thm]{Definition}
\newtheorem{rem}[thm]{Remark}
\newtheorem{prf}[thm]{Proof}

\maketitle


\begin{abstract}
 The problems of Hadamard quantum coin flipping in n-trials and related generalized Fibonacci sequences of numbers were introduced in \cite{1}.  It was shown that for an arbitrary number of repeated consecutive states, probabilities are determined by Fibonacci numbers for duplicated states, Tribonacci numbers for triplicated states and N-Bonacci numbers for arbitrary N-plicated states. In the present paper we generalize these results for direct product of multiple qubit states and arbitrary position of repeated states. The calculations are based on structure of Fibonacci trees in space of qubit states, growing in the left and in the right directions, and number of branches and allowed paths on the trees. By using $n$-qubit quantum coins as random n-qubit states with maximal Shannon entropy, we show that quantum  probabilities can be calculated by means of geometric probabilities. It illustrates possible application of geometric probabilities in quantum information theory. The Golden ratio of probabilities and the limit of n going to infinity are discussed.
\end{abstract}

\section{Introduction}
   
Fibonacci Numbers, 1,1,2,3,5,8,13,... satisfy recursion formula
 $F_n = F_{n-1} + F_{n-2}$, with initial conditions $F_1 = F_2 = 1$
and are related with Golden ratio $\varphi= \frac{1 + \sqrt{5}}{2}$ $\approx 1.6$, $\varphi' = -1/\varphi$, by the
Binet formula
 $$ F_n = \frac{\varphi^n - {\varphi'}^n}{\varphi - \varphi'}.$$
By interpreting this formula as a specific $q$-number in the post-quantum or $pq$-calculus, the quantum calculus of Fibonacci numbers (the Golden calculus) was introduced  in \cite{2}, and then generalized to Fibonacci divisors in \cite{3}. Relations of Fibonacci numbers with quantum measurement problem of qubits have been studied in \cite{1}. It was shown that probabilities  of measurement for qubit states with repeated identical states in $n$ trials are related with Fibonacci, Tribonacci and in general, $N$-Bonacci numbers.
This problem, as a quantum coin tossing problem, can be regarded as a quantum analogue of the classical coin tossing problem, apparently first introduced by A. de Moivre in 18th century,
in his book
\cite{m}. For relation of this problem with classical dynamical systems see \cite{k}.

In this work we generalize these results for the direct product of multiple qubit states and for arbitrary position of repeated states. The calculations are based on structure of Fibonacci trees in space of qubit states, and number of branches and allowed paths on the trees in Fibonacci garden. By using tensor product of maximally random multiple qubit states, we naturally come  to problems of geometrical quantum probabilities.
Applications of geometrical concepts as distance and area to calculate probability distributions are known as geometric probabilities.
The oldest problem in geometric probability is Buffon's Needle problem (1777), determining probability of needle crossing one of the lines on the page
and directly related to value of number $\pi$. In quantum information theory, the unit of quantum information is represented geometrically as
unit sphere, known as the
Bloch sphere. By stereographic projection of Bloch sphere (considered as the Riemann sphere) to complex plain , it is possible to calculate
some characteristics of qubits in terms of the plain geometry. This includes the qubit coherent states and the Apollonius representation \cite{t}.
As we show in the present paper, geometrical concepts can be used also to analyze quantum probabilities. First we show that
probabilities of measurement of one qubit state can be calculated in pure geometrical way by ratio of areas of spherical cap and the Bloch sphere itself. Then,
by using quantum n-qubit coins as maximally random n-qubit states, we show that the quantum measurement probabilities can be calculated by means of geometric probabilities. This illustrates possible application of geometric probabilities in quantum information theory.

\section{Geometric Quantum Probability on Bloch sphere}
The qubit  unit of quantum information is the state
 \begin{equation}
| \psi\rangle = c_0 |0\rangle + c_1 |1\rangle \in H = C^2,\label{qubit}
\end{equation}
where
$     |c_0|^2 + |c_1|^2    = 1$ and
\begin{equation}|0 \rangle = \left( \begin{array}{c} 1 \\ 0\end{array}  \right), \,\,\,\,|1 \rangle = \left( \begin{array}{c} 0 \\ 1\end{array}  \right).
\end{equation}
It can be parameterized
\begin{equation} |\theta,\varphi \rangle = \cos \frac{\theta}{2} |0 \rangle + \sin \frac{\theta}{2} e^{i\varphi} |1\rangle
\end{equation}
by points $(\theta, \varphi)$ on unit sphere $S^1$: $0 \le \theta \le \pi, 0 \le \varphi \le 2 \pi$.
Then, probabilities of the qubit measurement
\begin{equation}
p_0 = |c_0|^2 =\cos^2 \frac{\theta}{2} ,\hskip1.5cm  p_1 = |c_1|^2 = \sin^2 \frac{\theta}{2}, \hskip1.5cm p_0 + p_1 =1, \label{prob}
\end{equation}
are independent of angle $\varphi$ and geometrically are axial symmetry invariants. These probabilities have simple geometrical description in terms of
areas on Bloch sphere.
The area of spherical cap with radius $R$ (solid angle) is
 \begin{equation} A_\theta  = \int^\theta_0 \int^{2\pi}_0 R^2 \sin \theta' d\theta' d\varphi = 2\pi R^2 (1-\cos \theta) \end{equation}
or
\begin{equation}A_\theta = 4\pi R^2 \sin^2 \frac{\theta}{2}.   \end{equation}
Comparing with total area of the sphere $A = 4\pi R^2$, we get
pure geometric form of quantum probabilities   (\ref{prob}), as relative area of the spherical cap
\begin{equation} p_0 = \sin^2 \frac{\theta}{2}    = \frac{A_\theta}{A} \end{equation}
and the complementary  area
\begin{equation}p_1 = \cos^2 \frac{\theta}{2}    = \frac{A - A_\theta}{A}.  \end{equation} 
 
\section{n-qubit State as Random Variable}
Every binary number,
\begin{equation} N =  \sum^{n-1}_{k=0}    a_k 2^k = a_{n-1}a_{n-2}...a_1 a_0 ,             \end{equation}
where $a_k = 0,1$, determines $n$-qubit state in binary form
\begin{equation}  |N\rangle_n = |  a_{n-1}a_{n-2}...a_1 a_0  \rangle = |a_{n-1}\rangle \otimes    |a_{n-2}\rangle \otimes ... \otimes |a_1\rangle \otimes  |a_0\rangle . \end{equation}
This state can be generated by one qubit flipping gate $X = \sigma_1$,
in the form
\begin{equation} |N\rangle_n = |  a_{n-1}a_{n-2}...a_1 a_0  \rangle = \sigma_1^{a_{n-1}}\otimes  \sigma_1^{a_{n-2}}\otimes...\otimes\sigma_1^{a_{1}}\otimes \sigma_1^{a_{0}} |0\rangle_n ,\end{equation}
where
$\sigma_1 |0\rangle = |1\rangle,\,\,\,\,\,\sigma_1 |1\rangle = |0\rangle $. Then,
for arbitrary $n$ we have the set of $2^n$ orthonormal states $|0\rangle, |1\rangle, |2\rangle,...,|2^n-1\rangle$, as computational basis.
The normalized generic $n$-qubit state
 in this
computational basis is
 \begin{equation}|\psi \rangle = \sum^{2^n -1}_{k=0}      c_k |k\rangle ,  \hskip1.5cm \langle \psi|\psi\rangle = \sum^{2^n -1}_{k=0}  |c_k|^2 =1 . \label{n-qubit}  \end{equation}
According to the Born rule, the
measurement probability of collapse to basis state $ |k\rangle   $ is
$p_k = |c_k|^2$ and $\sum^{2^n-1}_{k=0} p_k =1$. This shows that state (\ref{n-qubit}) can be considered as a random variable state, with $2^n$ output states, as computational basis
states and the corresponding probabilities. Then,
the level of randomness for this state in form of the Shannon entropy is
\begin{equation} S =- \sum^{2^n -1}_{k=0} p_k   \log     p_k  = - \sum^{2^n -1}_{k=0} |c_k|^2   \log     |c_k|^2. \label{entropy} \end{equation}
For maximally random n-qubit state, the entropy is maximal $S = S_{max}$ and probabilities
 are equal and independent of $k$, $p_k = |c_k|^2 = 1/2^n$, so that
$ c_k = e^{i \phi_k}/\sqrt{2^n}. $
 Then, up to global phase, the maximally random states have the form
\begin{equation}  |\psi_{max}   \rangle = \frac{1}{\sqrt{2^n}}  \left(  |0\rangle + e^{i\varphi_1}  |1\rangle +...+  e^{i\varphi_{2^n-1}}  |2^n-1\rangle   \right). \label{max}\end{equation}
\begin{prop} The maximal value of Shannon entropy for $n$-qubit state is equal to number of qubits
\begin{equation} S_{max}   = n.             \end{equation}
For an arbitrary state, the entropy is bounded by inequality $0 \le S \le n$ and it is maximal for states (\ref{max}).
\end{prop}
\begin{prop}
 For separable normalized states
\begin{equation} |\Psi \rangle = |\psi \rangle_n \otimes | \phi \rangle_m    \end{equation}
the Shannon entropy is additive
\begin{equation}   S = S_n + S_m.  \label{n+m}\end{equation}
\end{prop}
Indeed, if
\begin{equation} |\Psi \rangle = \sum_{i_1,...,i_{n+m}=0,1}    c_{i_1,...,i_{n+m}} |i_1,...,i_{n+m} \rangle   \end{equation}
and
\begin{equation}|\psi \rangle_n = \sum_{i_1,...,i_{n}=0,1}    a_{i_1,...,i_{n}} |i_1,...,i_{n} \rangle,\,\,\,\,
|\phi \rangle_m = \sum_{j_1,...,j_{m}=0,1}    b_{j_1,...,j_{m}} |j_1,...,j_{m} \rangle, \label{2states} \end{equation}
then
$  c_{i_1,...,i_{n+m}} =     a_{i_1,...,i_{n}}  b_{i_{n+1},...,i_{n+m}}                     $
and probabilities
\begin{equation}  p_{i_1,...,i_{n+m}}    =     |c_{i_1,...,i_{n+m}} |^2 =       |a_{i_1,...,i_{n}} |^2 | b_{i_{n+1},...,i_{n+m}}  |^2. \end{equation}
By substituting to entropy formula (\ref{entropy}) and using normalization conditions for states (\ref{2states}), we get (\ref{n+m}).
\begin{cor}
 If
\begin{equation} S \neq S_n + S_m \end{equation}
the state is not separable, it is entangled state.
\end{cor}

\begin{cor}
To every ordered partition of integer numbers
\begin{equation}  n = n_1 +...+n_{N}        \end{equation}
exists separable n-qubit state
\begin{equation}  |\Psi\rangle_n = |\psi_1\rangle_{n_1} \otimes.... \otimes|\psi_N\rangle_{n_N}  \end{equation}
\end{cor}
\subsection{Maximally random separable states}
\begin{cor}
For separable n-qubit state with partition $n = n_1 +...+n_{N}$,
\begin{equation}  |\Psi\rangle_n = |\psi_1\rangle_{n_1} \otimes.... \otimes|\psi_N\rangle_{n_N} \end{equation}
entropy is the addition
\begin{equation} S_n = S_{n_1} + ... S_{n_N} \end{equation}
\end{cor}
\begin{prop}
 The product of maximally random states is the n-qubit state, which is maximally random and
corresponding entropy $S = n$ is given by ordered partition to the related partial entropies $S_{n_k} = n_k$, $k=1,...,N$,
\begin{equation}  n = n_1 +...+n_{N}  \end{equation}
\end{prop}

 \subsubsection{Geometric quantum probability in 1D and maximally random $n$ qubit states }

 \begin{prop}   For maximally random state
 \begin{equation}  |\psi_{max}   \rangle = \frac{1}{\sqrt{2^n}}  \left(  |0\rangle + e^{i\varphi_1}  |1\rangle +...+  e^{i\varphi_{2^n-1}}  |2^n-1\rangle   \right) \end{equation}
probability of collapse to every basis state $|s\rangle$ is the same and equal $p_s = 1/2^n$.
Then, probability to measure state $|s\rangle$ in interval $k\le s< l$
is given by geometric ratio of intervals (proportional to number of states)
\begin{equation}  p_{k,l} = \frac{L_{k,l}}{L}   =   \frac{N_{k,l}}{N}    = \frac{l-k}{2^n} .           \end{equation}
\end{prop}
\begin{prf}
 The average of projection operator $\hat P_s = |s \rangle \langle s|$ gives probability $p_s = \langle \Psi | \hat P_s | \Psi \rangle $ of collapse
$| \Psi \rangle \rightarrow |s \rangle$.
For maximally random n-qubit state (\ref{max}) it gives
\begin{equation}  p_s = \langle \Psi | \hat P_s | \Psi \rangle = \frac{1}{2^n} .\end{equation}
Projection operator to subspace $H_{k,l} = \prod^{l-1}_{s=k} H_s\in H_{2^n}$ in Hilbert space is
\begin{equation} \hat P_{k,l}  =   |k \rangle \langle k|    +  |k+1 \rangle \langle k+1| + ... +   |l-1 \rangle \langle l-1|                \end{equation}
and probability of collapse to this subspace is
\begin{equation}    p_{k,l} = \langle \Psi | \hat P_{k,l} | \Psi \rangle =\underbrace{\frac{1}{2^n} + \frac{1}{2^n} +... +\frac{1}{2^n}}_{l-k}  = \frac{l-k}{2^n}   \end{equation}
\end{prf}

 \subsubsection{Geometric quantum probability in 2D and maximally random bipartite states}
 \begin{prop}
For maximally random bipartite state
 \begin{equation}  |\Psi_{max}   \rangle_{n+m} =  |\psi_{max}   \rangle_n \otimes |\phi_{{max}}   \rangle_m \end{equation}
probability of collapse to every basis state $|i\rangle_n |j\rangle_m $ is the same and equal $p = \frac{1}{2^{n+m}}$.
Then, probability to measure state $|i\rangle |j\rangle$ in interval $k_1\le i< l_1$,
$k_2\le j< l_2$
is given by ratio of rectangular areas
\begin{equation}  p_{k_1l_1,k_2l_2} = \frac{A_{k_1l_1,k_2l_2}}{A}   =   \frac{N_{k_1l_1,k_2l_2}}{N}    = \frac{(l_1-k_1)(l_2-k_2)}{2^{n+m}}   \end{equation}
\end{prop}
 
 \subsubsection{Geometric quantum probability in  arbitrary dimensions and maximally random multi-partite states}
\begin{prop}
For maximally random multi-partite state
 \begin{equation}|\Psi_{max}\rangle_n = |\psi_{1_{max}}\rangle_{n_1} \otimes.... \otimes|\psi_{N_{max}}\rangle_{n_N}     \end{equation}
probability of collapse to every basis state $|i_1\rangle_{n_{1}}... |i_{N}\rangle_{n_N} $, $n=n_1 +...+n_N$,
is the same and equal $p = \frac{1}{2^{n_1+...+n_N}}$.
Then, probability to measure state $|i_1\rangle ...|i_N\rangle$ in intervals
$k_1\le i_1< l_1,...,
k_N\le i_N< l_N$
is given by ratio of volumes of parallelepipeds
\begin{equation}  p_{k_1l_1,...,k_N l_N} = \frac{V_{k_1 l_1,...,k_n l_N}}{V}   =   \frac{N_{k_1l_1,...,k_N l_N}}{N}    = \frac{(l_1-k_1)...(l_N-k_N)}{2^{n_1+...+n_N}} \end{equation}
\end{prop}
 
\section{Quantum Coin and Maximally Random Qubit State}
For one qubit state (\ref{qubit})
the Shannon entropy (\ref{entropy}),
as measure of uncertainty in result of measurement \cite{d}, is maximal
 $S=1$  for $p_0 = p_1 = \frac{1}{2}$,    like for classical coin.  But now it gives $|c_0| = |c_1| = \frac{1}{\sqrt{2}}$ and the Hadamard type qubit states (the Quantum Coin States),
\begin{equation}
| \varphi\rangle = \frac{1}{\sqrt{2}} (|0\rangle + e^{i\varphi} |1\rangle).\label{quantumcoin}
\end{equation}
These states are generated by Hadamard gate $H$ and the phase-shift gate $R_z(\varphi)$:  $|\varphi\rangle = R_z(\varphi) H |0\rangle$.
For $\varphi = 0$ and $\varphi = \pi$ we have the Hadamard states $|+\rangle$ and $|-\rangle$, correspondingly.
Flipping of the quantum coin is an  application of $X$ gate on "heads" $|0\rangle$
and "tails"  $|1\rangle$ states:
$  X |0\rangle = |1\rangle$ , $      X |1\rangle = |0\rangle .               $
Then,  applying the Hadamard gate  to the coin, initialized in state $|0\rangle$, quantum computer produces state
 $|+ \rangle = H |0\rangle$.
The measurement $M$ of this quantum coin state,
$$ |+\rangle = \frac{1}{\sqrt{2}}( |0\rangle +  |1\rangle)  \hskip 0.5cm \line(1,0){50}\fbox{\rule[-.3cm]{0cm}{1cm} M}  \line(1,0){50}  \hskip 0.5cm |i\rangle $$
gives states $|0\rangle$ or $|1\rangle$ with equal probabilities $p_0 = p_1 = \frac{1}{2}$. 
 
\section{Duplicated ${|1\rangle}$ States and Fibonacci Numbers}
The following problem is a quantum mechanical analog of classical coin flipping problem of de Moivre \cite{m}. The problem is to
find probability of measurement quantum coin states in $n-$ trials, such that repeated pattern of states $|1 \rangle$ appears only in last two final measurements.
The results of  these measurements we can order as n-qubit computational basis state.
Then, the first question is how many n-qubit states $A_n$ of following form exist:
\begin{equation}
\underbrace{{|*\rangle} \otimes { |*\rangle} \otimes  ...{ |*\rangle} }_{n-2}\otimes \underbrace{{ |1 \rangle} \otimes { |1 \rangle}}_{2}  \equiv \underbrace{{ |*\rangle |*\rangle   ... |*\rangle} { |1 \rangle  |1 \rangle}}_{n}  \label{duplicated}
\end{equation}

\subsection{Number of coin states and Fibonacci numbers}

To calculate number of allowed states, we notice that if state $|*\rangle = |0\rangle$, then preceding state from the left can be both, $|0\rangle$ or $|1\rangle$ state. But if
$|*\rangle = |1\rangle$ state, then the preceding state can only be $|0\rangle$. This implies the tree of states as Fibonacci tree, where $|0\rangle$ state plays the role of adult rabbit
and can produce young rabbit state. In contrast, state $|1\rangle$ corresponds to young rabbit  and can become adult  only.
The number of states $A_n$ in Fibonacci tree for $n$ qubit quantum states
 is equivalent to number of different paths (of length $n$ ) in this tree.
Starting from $n=3$ and  $|0\rangle$ state, number of paths satisfies the recursion formula
$
A_n = A_{n-1} + A_{n-2}
$
and initial conditions $A_2 = A_3 =1$.
This gives number of states as Fibonacci number
$
A_n = F_{n-1}, \,\,\,n = 2,3,...
$

\subsection{Quantum coin measurement probability}

 The measurement of quantum coin in $n$ trials gives states $|0\rangle $ and $|1\rangle $ with  probabilities $p_0 = p_1 = \frac{1}{2}$. This is equivalent to measurement
of maximally random $n$-qubit state (\ref{max}). If $\hat P_F$ is projection operator to allowed duplicated states (\ref{duplicated}), then
probability of collapse to corresponding subspace is
\begin{equation}
P_n =  \langle \psi_{max} | \hat P_F   | \psi_{max} \rangle =     A_n \,\frac{1}{2^n} = \frac{F_{n-1}}{2^n}, \,\,\,n = 2,4,....
\end{equation}
 Probabilities $P_n$ satisfy recursion formula for the Fibonacci polynomial numbers
 \begin{equation}
P_n = \frac{1}{2}P_{n-1} + \frac{1}{2^2}P_{n-2}, \label{PFibonaccirecursion}
\end{equation}
with initial values $P_1 = 0, P_2 = \frac{1}{2^2}$.
First few numbers are $P_3 = \frac{1}{2^3}$ , $P_4 = \frac{2}{2^4}$, $P_5 = \frac{3}{2^5}$, $P_6 = \frac{5}{2^6}$, etc.

\subsubsection{Golden Ratio and quantum probability}
If we compare the number  of duplicated $n$-qubit states $A_n = F_{n-1}$ and $n+1$-qubit states
$A_{n+1} = F_{n}$,
then
 \begin{equation} \lim_{n \rightarrow \infty} \frac{A_{n+1}}{A_n} = \lim_{n \rightarrow \infty} \frac{F_{n}}{F_{n-1}} = \varphi \end{equation}
   is the { Golden Ratio}. For corresponding probabilities we have half of the Golden Ratio
\begin{equation}\lim_{n \rightarrow \infty} \frac{P_{n+1}}{P_n} = \lim_{n \rightarrow \infty} \frac{1}{2}\frac{F_{n}}{F_{n-1}} = \frac{1}{2}\varphi. \end{equation}
\section{Arbitrary position of duplicated ${ |1\rangle}$ states}
If duplicated states appear at the end of $n$ qubit state, the
Fibonacci tree is growing to the left
 \begin{equation} { \leftarrow}
\underbrace{{|*\rangle} \otimes { |*\rangle} \otimes  ...{|*\rangle} }_{n-2}\otimes \underbrace{{|1 \rangle} \otimes { |1 \rangle}}_{2} ,\,\,\,\,\,A_n = F_{n-1},\end{equation}
while, for the states at the beginning, it is growing to the right
\begin{equation} { \rightarrow}
\underbrace{{ |1 \rangle} \otimes {|1 \rangle}}_{2} \otimes\underbrace{{|*\rangle} \otimes { |*\rangle} \otimes  ...{ |*\rangle} }_{n-2} ,\,\,\,\,\,A_n = F_{n-1}.\end{equation}
In generic case, when duplicated states appear at positions $k$ and $k+1$, where $k=1,2,...,n-1$, we have two Fibonacci trees, growing in both directions
 \begin{equation}
\underbrace{{|*\rangle} \otimes { |*\rangle} \otimes  ...{ |*\rangle} }_{k-1}\otimes \underbrace{{ |1 \rangle} \otimes {|1 \rangle}}_{k
\,\,\,\,\,\,\,\,\,\,\,\, k+1} \otimes\underbrace{{|*\rangle} \otimes { |*\rangle} \otimes  ...{|*\rangle} }_{n-k-1} \label{arbitrary}
\end{equation}
and the number of allowed states is product
\begin{equation} A_n = F_k \,\cdot\,F_{n-k} .\end{equation}

\subsection{Probability for arbitrary position of duplicated states}
Probability of states (\ref{arbitrary}) with arbitrary position of duplicated states is
 \begin{equation} P_{n,k} = \frac{F_{k} \cdot F_{n-k}}{2^n} = \frac{L_n - (-1)^k L_{n-2k}}{5\cdot 2^n},\end{equation}
where we have used following identity with Lucas numbers $L_n$:
\begin{equation}  F_m \cdot F_n = \frac{L_{m+n} - (-1)^n L_{m-n}}{5} . \end{equation}
 The number of states with pair ${ |1\rangle |1\rangle}$ in all possible positions $k=1,2,...,n-1$ is
\begin{equation}\sum^{n-1}_{k=1} F_k \cdot F_{n-k} = \frac{n L_n - F_n}{5}
\end{equation}
and probability of getting this pair anywhere, but only once is
\begin{equation} \sum^{n-1}_{k=1} P_{n,k}      = \sum^{n-1}_{k=1} \frac{F_k \cdot F_{n-k}}{2^n} = \frac{n L_n - F_n}{5\cdot 2^n}.
\end{equation}

\section{Fibonacci Numbers for Separable States}
Here we extend the above results to separable duplicated states.

\subsection{Fibonacci numbers for bipartite states}
For bipartite separable states
\begin{equation}
|\Psi \rangle_{n+m} =
\underbrace{|{* * ...*}{ 1 1}\rangle }_{n}\otimes  \underbrace{|{* * ...*}{ 1 1}\rangle }_{m}
 \end{equation}
the number of allowed duplicated states is
\begin{equation} A  = F_{n-1} F_{m-1} = \frac{1}{5}(L_{n+m-2} + (-1)^m L_{n-m}) \end{equation}
and corresponding probability is equal
\begin{equation} p = \frac{F_{n-1} F_{m-1}}{2^{n+m}} .                         \end{equation}

 \subsection{Fibonacci numbers for multi-partite states}
For multi-partite states of following form
  \begin{equation}
|\Psi \rangle_{n} =
\underbrace{|{ * * ...*}{ 1 1}\rangle }_{n_1}\otimes  \underbrace{|{ * * ...*}{ 1 1}\rangle }_{n_2}\otimes ... \otimes  \underbrace{|{* * ...*}{1 1}\rangle }_{n_N}  , \end{equation}
where partition $n=n_1 +...+n_N$,
the number of allowed duplicated states is
\begin{equation}A = F_{n_1-1}  F_{n_2-1}  ... F_{n_N-1}\end{equation}
and corresponding probability is equal
\begin{equation} p = \frac{F_{n_1-1} F_{n_2-1}... F_{n_N-1}}{2^{n_1+n_2+...+n_N}} .                         \end{equation}
\subsection{Arbitrary positions in bipartite states and Fibonacci garden}
If we have bipartite $n+m$ qubit state of the form
 \begin{equation}
|\Psi \rangle_{n+m} =
\underbrace{|{ * * ...*}{ 1 1} { *...*}\rangle }_{n}\otimes  \underbrace{|{* * ...*}{ 1 1}{ *...*}\rangle }_{m}  ,
\end{equation}
where duplicated states $|1\rangle |1\rangle$ take place at positions $(k_1,k_1+1)$ in $n$ qubit state and  $(k_2,k_2+1)$ in $m$ qubit state, then
the number of allowed states is
\begin{equation}A = F_{k_1} F_{n-k_1} F_{k_2} F_{m-k_2}  \end{equation}
and corresponding probability equal to
\begin{equation} p = \frac{F_{k_1} F_{n-k_1} F_{k_2} F_{m-k_2}}{2^{n+m}} .      \end{equation}

 \subsection{Arbitrary positions in multipartite states}
For separable $n$ qubit state with partition $n=n_1 +...+n_N$, and arbitrary positions of duplicated states, $(k_s, k_{s+1})$ in corresponding $n_s$ qubit state
($s = 1,2,...,N$),
 \begin{equation}
|\Psi \rangle_{n} =
\underbrace{|{ * * ...*}{ 1 1}{ *...*}\rangle }_{n_1}\otimes  \underbrace{|{ * * ...*}{ 1 1}{*...*}\rangle }_{n_2}\otimes ... \otimes
\underbrace{|{ * * ...*}{ 1 1} { *...*}\rangle }_{n_N}
\nonumber \end{equation}
the number of states is
\begin{equation}A = F_{k_1}F_{n_1-k_1} F_{k_2} F_{n_2-k_2}  ... F_{k_N} F_{n_N-k_N}\end{equation}
and corresponding probability is equal
\begin{equation} p = \frac{F_{k_1}F_{n_1-1} F_{k_2}F _{n_2-1}... F_{k_N} F_{n_N-1}}{2^{n_1+n_2+...+n_N}} .                  \end{equation}

 \section{Maximally  Random n-qubit State as Qudit Coin}
From computational $n$ qubit states $|k\rangle$, $k =0,1,2,...,2^n-1 $, the Hadamard gate in $2^n$ dimensions
\begin{equation} H = \frac{1}{\sqrt{2^n}}  \sum^{2^{n-1}}_{k,l=0}    \bar q^{k\cdot l} |k\rangle\langle l| ,  \end{equation}
where $q = e^{i \frac{2\pi}{2^n}}$ is primitive root of unity $q^{2^n} =1$,  can generate
$2^n$ maximally random  $n$-qubit states
\begin{equation}   |\phi_k \rangle = H |k\rangle,\,\,\,\,k=0,1,...,2^{n-1}.  \end{equation}
	By using Silvester shift matrix $\Sigma_1$, computational states are expressible as	
	\begin{equation} |k\rangle = \Sigma^k_1 |0\rangle \end{equation}
	and as follows
	\begin{equation} |\phi_k \rangle = \frac{1}{\sqrt{2^n}} [2^n]_{\bar q^k \Sigma_1} |0\rangle = A^+_k |0\rangle, \label{phi}\end{equation}
	where we have used following definition.
	\begin{definition} The matrix Q-number is defined by the sum
	$$  I + \Sigma_1 +  \Sigma^2_1   + ... + \Sigma_1^{2^n-1}    \equiv     [2^n]_{\bar \Sigma_1}.                 $$
\end{definition}
In Eq. (\ref{phi}), for every $k$ we have this matrix for $Q = \bar q^k \Sigma_1$, $k=0,1,2,...,2^n-1$.

Every maximally random $n$ qubit state $|\phi_k \rangle$ represents a qudit quantum coin with number of states $d = 2^n$ and the number of such coins is equal $2^n$.
The set of these quantum coin states is orthonormal and complete
\begin{equation}  \langle \phi_k | \phi_l \rangle = \delta_{kl}, \hskip1.5cm \sum^{2^n-1}_{k=0}      |\phi_k \rangle \langle \phi_k | = I.            \end{equation}
 
\subsection{n-qubit coin in M trials }
 For every quantum coin $|\phi_k \rangle$ the result of measurement is one of the states $|l\rangle$, $l=0,1,...,2^{n-1}$, with equal probability
\begin{equation}p = \frac{1}{2^n} = |\langle l |\phi_k \rangle|^2 .\end{equation}
This is why the state $|\phi_k \rangle$ represents the qudit coin with $d = 2^n$ states.
The measurement of an arbitrary qudit coin in $M$ trials, for duplicated states of the form
\begin{equation}\underbrace{{|*\rangle_n} \otimes { |*\rangle_n} \otimes  ...{ |*\rangle_n} }_{M-2}\otimes \underbrace{{ |1 \rangle_n} \otimes { |1 \rangle_n}}_{2}  \equiv \underbrace{{ |*\rangle |*\rangle   ... |*\rangle} { |1 \rangle  |1 \rangle}}_{M}
\end{equation}
was described in \cite{1}. Applying these results to our problem, we find that
number of allowed states $A_M = D_{M-1}$ is determined by generalized Fibonacci numbers $D_M$, with recursion formula
\begin{equation}  D_M = (2^n-1)  (D_{M-1} + D_{M-2}) ,\,\,\,\,D_0 =0, \,D_1 =1.  \end{equation}
Corresponding
probability for allowed states is equal
\begin{equation} P_M = \frac{D_{M-1}}{(2^n)^M} \end{equation}
and it satisfies the recursion relations
\begin{eqnarray} P_{M+1}  & =& \left(1- 2^{-n}  \right) \left(  P_M + 2^{-n} P_{M-1}    \right), \\  P_2 &=& 2^{-2n},\,\,\,P_3 = 2^{-2n} - 2^{-3n}.\end{eqnarray}

\subsection{Arbitrary n-qubit state in M trials}
The result can be generalized to generic $n$ qubit state
 \begin{equation} |\psi\rangle = \sum^{2^n-1}_{k=0} c_k |k\rangle_n \end{equation}
with probabilities to collapse
\begin{equation}p_k = |c_k|^2 = |\langle k |\psi \rangle|^2, \hskip1.5cm  \sum_{k=1}^{2^n-1} p_k =1.\end{equation}
For such state, considered as generic (not maximally random) $d = 2^n$ coin, probability of collapse for duplicated states $|1\rangle |1\rangle$ in M trials is determined by
recursion relations
\begin{eqnarray}P_M & = &(1-p_1) (P_{M-1} + p_1 P_{M-2}), \\
P_2 &=& p_1^2,\,\,\,P_3 = p_1^2 (1-p_1). \end{eqnarray}

\end{document}